\documentclass[11pt,a4paper]{article}

\usepackage[T1]{fontenc}
\usepackage[utf8]{inputenc}
\usepackage{lmodern}
\usepackage{amsmath,amssymb,mathtools,bm}
\usepackage{graphicx}
\usepackage{booktabs}
\usepackage{geometry}
\usepackage{hyperref}
\usepackage{microtype}

\newcommand{\dd}{\mathrm{d}}

\newcommand{\Mfun}{\mathcal{M}}
\newcommand{\Ham}{\mathcal{H}}
\newcommand{\KHam}{\mathcal{K}}
\newcommand{\Deltar}{\Delta_{r}}
\newcommand{\Deltat}{\Delta_{\theta}}
\newcommand{\Sig}{\Sigma}
\newcommand{\Lam}{\Lambda}
\newcommand{\Om}{\Omega}
\newcommand{\scrR}{\mathcal{R}}
\newcommand{\scrT}{\mathcal{T}}

\title{Observational Features of Nonsingular Rotating Black Holes in the
Dark-Energy Dominated Universe}

\date{}

\begin{document}

\author{Ram\'{o}n Torres\\
\small Dept. de F\'{i}sica, Universitat Polit\`{e}cnica de Catalunya, Barcelona, Spain.\\
\small \texttt{E-mail: ramon.torres-herrera@upc.edu} }

\maketitle

\begin{abstract}
Recent DESI baryon acoustic oscillation measurements, combined
with cosmic microwave background and type-Ia supernova data,
suggest that dark energy may evolve with cosmic time.
Along a cosmological background, an evolving dark-energy density
can be expressed locally as a function of a monotonic curvature
invariant, motivating an effective curvature-dependent
cosmological term.
Moreover, quantum field theory in curved spacetime and approaches to
quantum gravity with running gravitational couplings provide
complementary motivations for exploring such a term.

Here we study how the phenomenology of nonsingular rotating
black holes embedded in a Universe dominated by dark energy
can differ substantially from that of the Kerr and
Kerr--de Sitter geometries. We carry out the analysis in a stationary and axisymmetric spacetime that is characterized by an effective mass
profile and a running cosmological term.
The problem is directly formulated in phase space using a null geodesic
super-Hamiltonian and a Mino-type parameter.  
The shadows of the black holes are calculated
by backward ray tracing from a zero-angular-momentum observer.
We define and analyze a comprehensive set of standard shadow
observables, characterizing the size, shape, and displacement
of the shadow.
We show that the area of the shadow and its vertical diameter provide particularly clear signatures of a running cosmological term.
In this way, joint analyses of multiple strong-field observables and of several black holes with different masses and spins could provide a realistic route to tightly constraining, or potentially measuring, the effective regularization scale behind the running of the cosmological term and statistically distinguishing the non-singular black hole geometry from Kerr/Kerr-dS solutions.

\end{abstract}

\section{Introduction}

Recent cosmological observations have renewed interest in scenarios in which the dark-energy sector is not strictly constant. In particular, the first DESI baryon-acoustic-oscillation (BAO) measurements, when combined with cosmic-microwave-background (CMB) and type-Ia supernova data, showed a preference for an evolving dark-energy equation of state within commonly used two-parameter descriptions \cite{DESI2024}. This trend persisted and became more significant with the DESI Data Release 2 (DR2) BAO measurements \cite{DESIDR2}, and was found to be stable under a variety of parametric and non-parametric reconstructions of the dark-energy sector \cite{Lodha2025}. The most recent DESI Ly$\alpha$ full-shape analysis somewhat reduces the significance of the departure from $\Lambda$CDM, while still leaving a preference for evolving dark energy when DESI is combined with CMB and supernova observations \cite{DESILya2026}. Thus, although a cosmological constant remains compatible with the available data and the evidence cannot yet be regarded as conclusive, current observations provide non-negligible motivation for considering an effective dark-energy component that evolves during the cosmological expansion. Since the curvature invariants of a homogeneous FLRW universe are themselves functions of cosmic time, such an evolution may equivalently be viewed
as an effective dependence of the dark-energy sector on the characteristic curvature scale of the Universe.

Moreover, quantum field theory in curved spacetime provides an
independent motivation for a cosmological sector sensitive to
the background geometry. Explicit running-vacuum calculations
in FLRW spacetime yield a renormalized vacuum energy density
that depends on the Hubble rate and its time derivatives,
supporting an effective description in terms of evolving
geometric scales~\cite{ShapiroSola2009,MorenoPulidoEtAl2023}.

Scale dependence also arises in renormalization-group
approaches to quantum gravity. In the asymptotic-safety programme, the gravitational
effective action evolves with a coarse-graining scale $k$; within the
Einstein--Hilbert truncation, this evolution is encoded in running
couplings $G(k)$ and $\Lambda(k)$~\cite{Reuter1998}. Connecting this scale
dependence to a particular spacetime requires a physical prescription for
identifying $k$ with scales determined by the geometry and the effective
dynamics~\cite{BorissovaPlatania2023}.
A curvature-based identification of $k$ then induces an
effective curvature dependence in both couplings.

In renormalization-group-improved black-hole models,
the ultraviolet behaviour of both $G(k)$ and $\Lambda(k)$
can determine whether the classical singularity
persists~\cite{AdeifeobaEtAl2018}.
These results motivate investigating how a running cosmological
term affects the geometry and observable properties of
nonsingular black holes.

The classical Kerr and Kerr--de Sitter solutions provide
the reference geometries for assessing these effects,
with vanishing and positive cosmological constant,
respectively.
The shadows of these reference geometries have been studied
analytically~\cite{Bardeen1973,GrenzebachPerlickLaemmerzahl2014}.
For a strictly constant cosmological term at the value associated
with the present dark-energy density,
$\Lambda\simeq10^{-52}\,\mathrm{m}^{-2}$, its direct influence
on the photon region is exceedingly small. The relevant
dimensionless parameter is $\Lambda r_g^2$, where $r_g$ is
the gravitational radius; this parameter is only of order
$10^{-26}$ even for a black hole of mass $10^{10}M_\odot$.
For nearby sources, and at fixed mass, spin, and observer
configuration, the Kerr and Kerr--de Sitter shadows are
therefore practically indistinguishable at present imaging
precision~\cite{OmwoyoEtAl2022}. At cosmological distances,
expansion and observer motion can nevertheless affect the
apparent angular size and must be accounted for
separately~\cite{LiGuoChen2020}. This motivates investigating
whether a curvature dependent cosmological term can
produce appreciable shadow modifications while retaining
its small value on cosmological scales.

These reference geometries also retain the classical ring
curvature singularity: a constant positive $\Lambda$
changes the large-scale geometry and horizon structure
without removing it~\cite{Carter1968,LakeZannias2015}.
Regular black-hole models address this limitation through
effective nonsingular geometries, providing a framework
for studying photon propagation beyond the classical
solutions.
Representative constructions include the spherically symmetric
models without~\cite{Hayward2006} or with cosmological constant~\cite{TorresUnimodular2017}, rotating generalizations of the Bardeen and Hayward geometries~\cite{BambiModesto2013},
quantum-improved rotating black holes~\cite{Torres2017} and rotating models admitting a constant cosmological
term~\cite{NevesSaa2014}.
These constructions demonstrate that a suitable modification of the effective
mass profile (or, equivalently, a running-G) can remove scalar curvature divergences while preserving
a classical large-distance limit.

The family introduced in
Ref.~\cite{Torres2026} extends this setting by allowing both an effective
mass profile 
and a running cosmological term
$\Lambda$ within a generalized Kerr--Schild construction.
It includes Schwarzschild, Schwarzschild--de Sitter, Kerr
and Kerr--de Sitter as special cases, while suitable
profiles yield geometries without scalar polynomial
curvature singularities~\cite{Torres2026,TorresFayos2017}.

For shadow observations, the relevant question is whether
the modifications that regularize the interior also
affect photon trapping in the exterior. Horizon-scale imaging has
already been used to constrain asymptotic-safety-inspired
departures from Kerr in models without an explicit cosmological
term~\cite{HeldGoldEichhorn2019}. A running cosmological term can modify the shadow size and
shape through its interplay with the running mass profile.
We isolate these contributions by comparing the classical
reference geometry with configurations in which only
$\Lambda$ runs or both profiles run.

These shadow tests probe a curvature regime far beyond
that sampled by the cosmological reconstruction.
For the spatially flat FLRW background described in
Appendix~\ref{app:LambdaK}, the present value of the
Kretschmann scalar is $ K_0 \simeq 3.8\times10^{-104}\ {\rm m}^{-4}$,
whereas the highest-redshift DESI Ly$\alpha$ measurement entering the
current dark-energy analysis, at $z_{\rm eff}=2.33$
\cite{DESILya2026}, corresponds, for the central cosmological fit, to
approximately $K(z_{\rm eff})\simeq 5.8\times10^{-102}\ {\rm m}^{-4}$.
Thus, the cosmological reconstruction 
samples approximately the range $K\sim10^{-104}$--$10^{-102}\ {\rm
m}^{-4}$. 

Black-hole photon regions probe a vastly higher curvature regime.
For Sgr~A$^*$, with a mass of approximately
$4\times10^6\,M_\odot$~\cite{EHTSgrA2022}, the Schwarzschild
benchmark at $r_{\rm ph}=3M$ gives
$K_{\rm ph}=16/(243M^4)\simeq5\times10^{-41}\,\mathrm{m}^{-4}$,
illustrating the curvature scales relevant to horizon-scale
observations.
Although rotation modifies the curvature distribution,
this benchmark indicates a separation of roughly
sixty orders of magnitude between the two regimes
\footnote{Since at a fixed
dimensionless radius $r/M$ the black-hole curvature scales approximately
as $K\propto M^{-4}$, lower-mass black holes probe still substantially
larger curvature scales.}.

This enormous separation of curvature scales makes black-hole
observations complementary to cosmological probes. If the effective
cosmological term follows a universal curvature-dependent law, the two
regimes could constrain distinct portions of that law. In particular,
measurements spanning black holes of different masses and dimensionless
spins could probe a broad range of photon-region curvatures and thereby
extend the information available from low-curvature cosmological
observations. 
Nevertheless, extracting constraints from horizon-scale observations
requires accounting for degeneracies involving the black-hole mass and
spin, the viewing geometry, and the properties of the emitting
plasma~\cite{EHTM87VI2019,EHTSgrAVI2022}.

The aim of this work is to determine how a running cosmological term
modifies photon trapping and the shadows of nonsingular rotating black
holes, and to identify the observables most sensitive to these changes.
We construct black-hole shadows by backward ray tracing from
a zero-angular-momentum observer, using a Hamiltonian formulation
of null geodesic motion that does not require Hamilton--Jacobi
separability or a Carter constant. Using an explicit asymptotic-safety-inspired model,
we compare shadow sizes and morphologies with their classical Kerr
and Kerr--de Sitter counterparts under matched observer prescriptions.
Particular attention is paid to disentangling the effects of the
running mass and cosmological profiles, assessing the influence of
the viewing inclination, and determining when the effective regularization scale
produces appreciable changes in photon trapping relative to
the classical reference geometries. These comparisons provide a quantitative basis for assessing
how strong-field observations could constrain the model and complement
cosmological investigations of the vacuum sector.

The article is organized as follows. Section~\ref{secGeom} introduces the spacetime
geometry and the specific effective model considered in this work.
Section~\ref{secNullGeod} develops the Hamiltonian formulation of null geodesic motion
and examines how a running cosmological term affects separability.
Section~\ref{secTrappedGeod} examines photon trapping,
establishes an analytical obstruction to nonequatorial
spherical null geodesics, illustrates nonspherical
trapping, and analyzes equatorial light rings. Sections~\ref{secZAMO}--\ref{secObservables} introduce the zero-angular-momentum observer and local screen, describe the shadow construction, and define the
observables used to characterize its size and morphology. Section~\ref{secResults} presents validation against
classical limits, the expansion for small regularization
scale, comparisons isolating the running profiles,
and scans over regularization scale, spin and observer
inclination.
Section~\ref{secConclu} summarizes the findings and discusses prospects for
constraining the effective geometry through complementary observations.
Appendix~\ref{app:LambdaK} derives the present-epoch cosmological reconstruction of
$\Lambda(K)$, while Appendix~\ref{app:numerics} details the numerical ray-tracing
procedure and the checks required for reproducibility.

\section{Geometry of nonsingular rotating black holes embedded in Dark Energy}
\label{secGeom}

The family of nonsingular rotating black-hole geometries embedded in
a dark-energy background was constructed and analysed in
Ref.~\cite{Torres2026}. Here we consider the subclass admitting the
Boyer--Lindquist-like representation given below, in coordinates
$(t,r,\theta,\phi)$ adapted to the stationary and axial Killing fields
$\partial_t$ and $\partial_\phi$. For this subclass admitting a Boyer--Lindquist-like representation, the effective mass profile and cosmological term must depend
only on the radial coordinate~\cite{Torres2026}, $\mathcal{M}=\mathcal{M}(r)$ and $\Lambda=\Lambda(r)$.
The line element takes the form

\begin{align}
\dd s^2={}&
\Sig\left(
       \frac{\dd r^2}{\Deltar}
       +\frac{\dd\theta^2}{\Deltat}
     \right)
\nonumber\\
&+
\frac{\sin^2\theta\,\Deltat}{\Sig\,\Xi^2}
  \left[a\,\dd t-(r^2+a^2)\dd\phi\right]^2
-
\frac{\Deltar}{\Sig\,\Xi^2}
  \left[\dd t-a\sin^2\theta\,\dd\phi\right]^2 ,
\label{eq:metric}
\end{align}
where
\begin{align}
\Sig(r,\theta)&=r^2+a^2\cos^2\theta ,
\label{eq:Sigma}\\
\Deltar(r)&=
r^2+a^2-2r\Mfun(r)
-\frac{\Lam(r)}{3}r^2(r^2+a^2),
\label{eq:Deltar}\\
\Deltat(r,\theta)&=
1+\frac{a^2\Lam(r)}{3}\cos^2\theta ,
\label{eq:Deltatheta}\\
\Xi(r)&=1+\frac{a^2\Lam(r)}{3}
\label{eq:Xi}
\end{align}
and $a$ is the rotation parameter.  The spacetime is stationary and
axisymmetric.

The isolated roots of $\Deltar(r_H)=0$
are Killing horizons \cite{Torres2026}.  Following the terminology of
Ref.~\cite{Torres2026}, a Boyer--Lindquist block with
$\Deltar>0$ is a \textit{type-I block}, whereas $\Deltar<0$ defines a \textit{type-II
block}.  
For the asymptotically Kerr--de Sitter configurations considered
below, $\Lambda(r)\to\Lambda_0>0$ and
$\mathcal{M}(r)\to M>0$ as $r\to\infty$,
where $\Lambda_0$ and $M$ denote, respectively, the cosmological
constant and mass parameter of the asymptotic Kerr--de Sitter
geometry.
These limits imply
$\Delta_r(r)\sim-\Lambda_0 r^4/3$ as $r\to\infty$, so that
$\Delta_r$ is negative at sufficiently large $r$.
If an event horizon $r_+$ exists and $\Delta_r>0$ immediately
outside it, 
continuity therefore requires at least one further
zero at a larger radius. 
Let $r_{\mathrm{out}}:=\min\{r>r_+:\Delta_r(r)=0\}$.
The connected exterior type-I block adjacent to $r_+$ is therefore
$r_+<r<r_{\mathrm{out}}$.
For the configurations considered below, we restrict attention to
the case in which this outer boundary is the cosmological horizon,
$r_{\mathrm{out}}=r_C$. The stationary exterior domain relevant
for the observer is then $r_+<r<r_C$.

\subsection{Specific model}

In order to model the phenomenology of these black holes it is necessary to assume specific functions $\Mfun(r)$ and $\Lam(r)$. For the asymptotic-safety-inspired example introduced in
Ref.~\cite{Torres2026}, and adopting geometrized units
$G_0=c=1$, the functions are
\begin{equation}
\Mfun(r)=M\frac{r^3}{r^3+\ell^3},
\qquad
\Lam(r)=\frac{\ell^2+\Lambda_0 r^4}{\ell^4+r^4},
\end{equation}
where $\ell>0$ is the regularization scale, while $M$ and
$\Lambda_0$ are the limiting parameters defined above.
Note in particular that for $\ell>0$, if (and only if) $\Lambda_0\ell^2<1$ the cosmological profile $\Lambda(r)$ is strictly decreasing on $r\geq0$,
interpolating between $\Lambda(0)=1/\ell^2$ and
$\lim_{r\to\infty}\Lambda(r)=\Lambda_0$.

The particular choice above is not intended as a unique
first-principles prediction of asymptotically safe gravity, but as a
minimal effective realization of the ultraviolet and infrared
properties expected in that framework~\cite{Platania2023}.  Its motivation is
threefold.  First, the suppression
${\cal M}(r)\propto r^3$ at small $r$ implements the weakening of the
effective gravitational interaction suggested by gravitational
antiscreening \cite{Reuter1998,BonannoReuter2000,Platania2019} (this form was already implemented by Hayward in \cite{Hayward2006}), while the running cosmological term approaches a finite ultraviolet value with
$\Lambda(r)-\Lambda(0)=O(r^4)$. These behaviours are precisely 
required to avoid curvature singularities \cite{Torres2026}.  Second, both
functions recover their classical infrared limits,
${\cal M}(r)\rightarrow M$ and
$\Lambda(r)\rightarrow\Lambda_0$ as $r\rightarrow\infty$, so that the geometry continuously
approaches the standard Kerr--de Sitter solution outside the region
where the new scale $\ell$ becomes relevant.  Finally, the model is
deliberately economical: the single crossover scale $\ell$ controls
the transition between the regular ultraviolet core and the
classical exterior, while retaining closed analytic metric functions
that make the horizon structure, null dynamics and shadow
phenomenology directly computable.  
The profiles should therefore be regarded as simple interpolations
between the regular ultraviolet core and the classical Kerr--de Sitter
infrared regime, rather than as unique predictions for the intermediate
region. Smooth alternatives with the same limiting behaviour and a
comparable crossover scale may be expected to preserve the qualitative
phenomenology, while the quantitative shadow modifications remain
model dependent.

\section{Null geodesic dynamics}
\label{secNullGeod}

The shadow and photon-region structure of a black hole are determined
by the behaviour of null geodesics in its exterior geometry.  We
therefore will now formulate the photon dynamics for the metric introduced
above.

The Killing fields $\partial_t$ and $\partial_\phi$ imply the two
conserved canonical momenta
\begin{equation}
E\equiv-p_t,
\qquad
L\equiv p_\phi .
\label{eq:EL}
\end{equation}
For future-directed photons $E$ is the conserved Killing energy in a
region where $\partial_t$ has the usual asymptotic interpretation,
while $L$ is the conserved axial angular momentum.

The nonvanishing inverse-metric components required below are
\begin{align}
g^{rr}
&=\frac{\Deltar}{\Sig},\hspace{1cm}
g^{\theta\theta}
=\frac{\Deltat}{\Sig},
\label{eq:inverse1}\\
g^{tt}
&=
-\frac{\Xi^2 F}
       {\Sig\,\Deltar\,\Deltat},
\label{eq:gttinv}\\
g^{t\phi}
&=
-\frac{a\Xi^2
\left[(r^2+a^2)\Deltat-\Deltar\right]}
{\Sig\,\Deltar\,\Deltat},
\label{eq:gtphiinv}\\
g^{\phi\phi}
&=
\frac{\Xi^2
\left[\Deltar-a^2\Deltat\sin^2\theta\right]}
{\Sig\,\Deltar\,\Deltat\sin^2\theta},
\label{eq:gphiphiinv}
\end{align}

where
\begin{equation}
F(r,\theta)
=
(r^2+a^2)^2\Deltat
-a^2\sin^2\theta\,\Deltar .
\end{equation}

The geodesic Hamiltonian, also commonly referred to as the
geodesic super-Hamiltonian, is
\begin{equation}
\Ham(x^\mu,p_\mu)
=
\frac12 g^{\mu\nu}p_\mu p_\nu .
\label{eq:Hamiltonian}
\end{equation}
Null geodesics satisfy $\Ham=0$.
Hamiltonian formulations of Kerr geodesics and their separability go
back to Carter \cite{Carter1968}; see also
Refs.~\cite{Bardeen1973,Chandrasekhar1983}.
For computational convenience, we introduce the rescaled super-Hamiltonian
$\KHam\equiv \Sig\Ham$.
Substituting now the inverse metric components and the conserved
momenta into
$\KHam$ yields
\begin{align}
2\KHam={}&
\Deltar p_r^2
+\Deltat p_\theta^2
+\frac{\Xi^2}{\Deltat\sin^2\theta}
      \left(L-aE\sin^2\theta\right)^2
\nonumber\\
&-
\frac{\Xi^2}{\Deltar}
      \left[(r^2+a^2)E-aL\right]^2 .
\label{eq:KHam}
\end{align}
The null constraint is equivalently
\begin{equation}
\KHam=0.
\label{eq:Kzero}
\end{equation}

Let $\lambda_{\rm aff}$ denote an affine parameter for the original
Hamiltonian $\Ham$.  We introduce
\begin{equation}
\dd\gamma
=
\frac{\dd\lambda_{\rm aff}}{\Sig}.
\label{eq:minotime}
\end{equation}
On the null constraint surface $\Ham=0$, the Hamiltonian vector fields
generated by $\Ham$ and $\KHam=\Sig\Ham$ differ only by this
reparametrization.  Thus the same spacetime null geodesics are obtained
from $\KHam$, but parametrized by $\gamma$.
For Kerr geodesics, an analogous parameter was introduced by Mino
because it decouples the radial and polar motions
\cite{Mino2003}.  In the present running-$\Lambda$ geometry the motion
is generically not separable, so $\gamma$ should more precisely be
called a \emph{Mino-type parameter}.  Its usefulness remains: the
common factor $\Sig^{-1}$ is removed from the $(r,\theta)$ equations,
which substantially improves their form for numerical integration.

For compactness we define
\[
A\equiv (r^2+a^2)E-aL,
\qquad
B\equiv L-aE\sin^2\theta,
\]
\[
C\equiv
\frac{\Xi^2}{\Delta_\theta\sin^2\theta},
\qquad
D\equiv
\frac{\Xi^2}{\Delta_r}.
\]

With $E$ and $L$ held fixed, the Hamilton equations required
for the numerical construction of the shadow are
\begin{align}
\frac{\dd r}{\dd\gamma}
&=
\Deltar p_r,
\label{eq:rdot}\\
\frac{\dd\theta}{\dd\gamma}
&=
\Deltat p_\theta,
\label{eq:thetadot}\\
\frac{\dd p_r}{\dd\gamma}
&=
-\frac12
\left[
 \Deltar' p_r^2
 +(\partial_r\Deltat)p_\theta^2
 +(\partial_r C)B^2
 -(\partial_r D)A^2
 -2D A(2rE)
\right],
\label{eq:prdot}\\
\frac{\dd p_\theta}{\dd\gamma}
&=
-\frac12
\left[
 (\partial_\theta\Deltat)p_\theta^2
 +(\partial_\theta C)B^2
 +2CB(-2aE\sin\theta\cos\theta)
\right].
\label{eq:pthetadot}
\end{align}
Here primes denote differentiation with respect to $r$.
Equations~\eqref{eq:rdot}--\eqref{eq:pthetadot}, together with the
constants $E$ and $L$, form a closed first-order system.

The cyclic coordinates can, if required, be reconstructed from
\begin{equation}
\frac{\dd t}{\dd\gamma}
=
\Xi^2
\left[
\frac{(r^2+a^2)A}{\Deltar}
+
\frac{aB}{\Deltat}
\right]\ , 
\ \ \ 
\frac{\dd\phi}{\dd\gamma}
=
\Xi^2
\left[
\frac{aA}{\Deltar}
+
\frac{B}{\Deltat\sin^2\theta}
\right].
\end{equation}

\subsection{Separability and the role of a running cosmological term}

A central property of the metric is that the geodesic problem changes
qualitatively when $\Lam$ is promoted from a constant to a radial
function.
If $\Lam$ were effectively a constant $(\Lam'(r)=0)$,
then $\Xi$ would be a constant and $\Deltat$ would depend only on $\theta$.
Equation~\eqref{eq:KHam} separates into a radial and an angular part,
irrespective of the detailed radial form of $\Mfun(r)$.  One may
introduce a Carter-type separation constant $\mathcal{Q}$ through
\begin{equation}
\mathcal{Q}
=
\Deltat p_\theta^2
+
\frac{\Xi^2}{\Deltat\sin^2\theta}
\left(L-aE\sin^2\theta\right)^2 .
\label{eq:Carterlike}
\end{equation}
The equations then reduce to
\begin{align}
\left(\frac{\dd r}{\dd\gamma}\right)^2
&=
\scrR(r),
\label{eq:Rsep}\\
\left(\frac{\dd\theta}{\dd\gamma}\right)^2
&=
\scrT(\theta),
\label{eq:Tsep}
\end{align}
with
\begin{align}
\scrR(r)
&=
\Xi^2
\left[(r^2+a^2)E-aL\right]^2
-\Deltar\mathcal{Q},
\label{eq:Rpotential}\\
\scrT(\theta)
&=
\Deltat\mathcal{Q}
-
\frac{\Xi^2}{\sin^2\theta}
\left(L-aE\sin^2\theta\right)^2 .
\label{eq:Tpotential}
\end{align}
This is the structure underlying the familiar analytical
Kerr--(anti-)de Sitter photon-region calculation
\cite{Carter1968,GrenzebachPerlickLaemmerzahl2014,PerlickTsupko2022}.

However, for a genuinely running cosmological term,
$\Lam'(r)\neq0$, both $\Xi=\Xi(r)$ and
$\Deltat=\Deltat(r,\theta)$ contain radial dependence.
Consequently the expression on the right-hand side of
Eq.~\eqref{eq:Carterlike} is not, in general, a constant of motion.
The radial and polar sectors of Eq.~\eqref{eq:KHam} cannot be separated
by the usual Carter construction.
The calculation below therefore uses
Eqs.~\eqref{eq:rdot}--\eqref{eq:pthetadot} directly and makes no
separability assumption.  This is analogous to the numerical treatment
required in other stationary axisymmetric spacetimes with
nonintegrable null geodesic motion
\cite{CunhaPRL2015,CunhaPRD2016,CunhaFPO2017}.

\section{Trapped null geodesics and critical photon dynamics}
\label{secTrappedGeod}

Trapped null geodesics are null trajectories that remain confined to
a bounded region of the black-hole exterior instead of either crossing
the event horizon or escaping to the outer region. Their unstable
subset is of particular relevance to black-hole imaging, since nearby
null trajectories can spend a long time in the strong-field region
before eventually being captured or escaping. The critical dynamics
associated with this unstable trapped set determines the boundary
between captured and escaping rays and therefore underlies the
black-hole shadow observed on a local celestial sky
\cite{PerlickTsupko2022,CunhaFPO2017}.

For Kerr--de Sitter (where
$\mathcal M(r)=M$ and $\Lambda(r)=\Lambda_0$)
the exterior trapped null geodesics are precisely the
admissible \textit{spherical photon orbits},
$r=r_p=\mathrm{const}$. These satisfy
\[
\mathcal{R}(r_p)=0,
\qquad
\mathcal{R}'(r_p)=0,
\]
and generate the standard photon region of Kerr--de Sitter
spacetimes. Explicit expressions for the corresponding impact
parameters and the allowed polar range are given, for example, in
Refs.~\cite{GrenzebachPerlickLaemmerzahl2014,PerlickTsupko2022}.

However, for the rotating geometry~\eqref{eq:metric} with a
running cosmological term, the following proposition shows that,
under its hypotheses, spherical null geodesics are restricted
to equatorial light rings.

\medskip
\noindent\textbf{Proposition (Exterior spherical null geodesics).}
For the metric~\eqref{eq:metric}, consider a radius $r_p>0$
in a regular exterior region with
$\Delta_r>0$, $\Delta_\theta>0$, $\Xi>0$, and
$\mathcal M(r_p)\geq0$.
If $a\neq0$ and $\Lambda'(r_p)\neq0$, every nontrivial null
geodesic with $r=r_p$ is an equatorial light ring.

\smallskip
\noindent\emph{Proof.}
Constant radius requires $p_r=0$ and
$\dd p_r/\dd\gamma=0$ throughout the orbit. Using the previously defined $A$ and
$B$, introduce
\[
P=\frac{A^2}{\Delta_r},\qquad
J=L-aE,\qquad
\chi=aE-(\Xi-1)J,\qquad
\zeta=\frac{\cos^2\theta}{\Delta_\theta}.
\]
All radial quantities below are evaluated at $r_p$, with
radial derivatives taken at fixed $E,L$.
Eliminating $p_\theta^2$ between the null constraint
and radial equilibrium gives
\begin{equation}
\Xi P'
=\Xi'\left[\Xi P\zeta-2P+2(J+\chi\zeta)^2\right].
\label{eq:spherical_obstruction}
\end{equation}
If $\theta$ varies, $\zeta$ ranges over an interval, whereas
the left-hand side is constant. Since
$\Xi'=a^2\Lambda'/3\neq0$, the quadratic and linear
coefficients must vanish: 
$2\chi^2=0$, $\Xi P+4J\chi=0$.
These conditions imply $P=0$, contradicting the null constraint.
Hence $\theta$ must be constant.

It remains to exclude constant non-equatorial latitudes.
Away from the axis and equator, $p_\theta=0$, and angular
equilibrium requires
\[
2aE\,\Delta_\theta\sin^2\theta
+B\left[\Xi-2(\Xi-1)\sin^2\theta\right]=0.
\]
For a nonzero null covector, $B\neq0$.
Combining this equation with
$A=\Sigma E-aB$ and (from the null constraint)
$A^2=\Delta_r B^2/(\Delta_\theta\sin^2\theta)$ yields
\[
0=\Xi^2\Sigma^2
  +8a^2r_p\mathcal M(r_p)\Delta_\theta\sin^2\theta>0,
\]
a contradiction. 

To exclude the remaining subcase of a null geodesic at a fixed radius on the rotation axis, it is enough to notice that such a constant-radius curve would be timelike because $\Delta_r>0$.
Thus $\theta=\pi/2$, proving the claim.
\hfill$\square$

\smallskip
For the adopted profiles, $\mathcal M(r)>0$ and $\Lambda'(r)<0$
for every $r>0$ when $\ell>0$ and $\Lambda_0\ell^2<1$.
The proposition therefore applies throughout the exterior
of these rotating black holes: every trapped null geodesic with
polar motion necessarily has a varying radial coordinate.

\smallskip
\noindent\emph{Numerical example.}
For $a/M=0.8$, $\ell/M=0.83$, and
$\Lambda_0 M^2=10^{-6}$, numerical shooting gives an orbit
periodic in $(r,\theta,p_r,p_\theta)$, with
$L/(ME)\simeq2.930016273634$ and
$ME\,T_\gamma\simeq2.07444655$, where $T_\gamma$ is its
period in the Mino-type parameter $\gamma$.
The motion satisfies
\[
1.4450293734\lesssim r/M\lesssim1.4450582273,
\qquad
|\theta-\pi/2|\leq0.3.
\]
The one-period closure error in the dimensionless variables
$(r/M,\theta,p_r/E,p_\theta/(ME))$ is below
$3\times10^{-12}$. The resolved radial excursion,
$(r_{\max}-r_{\min})/M\simeq2.89\times10^{-5}$, illustrates
nonspherical trapping outside the event horizon,
$r_+/M\simeq1.34897418$.

\smallskip
Accordingly, we will construct the shadow by direct integration of
the full null geodesic equations, as described in
Sec.~\ref{sec:shadowconstruction}, while computing the equatorial
light rings separately below.

\subsection{Equatorial light rings for arbitrary \texorpdfstring{$\Lam(r)$}{L(r)}}

Under the hypotheses of the preceding proposition, equatorial
light rings are the only exterior spherical null geodesics.
We now determine their circular-orbit conditions and radial
stability. The analysis applies to arbitrary $\Lambda(r)$
and uses the invariant equatorial submanifold
$\theta=\pi/2$, $p_\theta=0$, on which
$\Delta_\theta=1$ and $\Sigma=r^2$.

For null geodesics with $E\neq0$, we introduce the impact parameter
$b=L/E$ and define the equatorial radial function
\begin{equation}
\scrR_{\mathrm{eq}}(r;b)
=
\left(r^2+a^2-ab\right)^2
-\Deltar(r)(b-a)^2.
\end{equation}
For a circular equatorial photon orbit in the black-hole exterior,
the null constraint and radial equilibrium imply, respectively,
\begin{equation}
\scrR_{\rm eq}(r_{\rm LR};b_{\rm LR})=0,
\qquad
\partial_r
\scrR_{\rm eq}(r_{\rm LR};b_{\rm LR})=0,
\label{eq:LRconditions}
\end{equation}
where the radial derivative is taken at fixed impact parameter $b$.

For all black-hole configurations considered below, the exterior equatorial light-ring equations yield two radially unstable solutions, corresponding to the prograde and retrograde branches,
\begin{equation}
(r_{\rm LR}^{-},b_{\rm LR}^{-}),
\qquad
(r_{\rm LR}^{+},b_{\rm LR}^{+}),
\label{eq:LRpm}
\end{equation}
with the assignment of the signs fixed by the chosen orientation
of $a$ and $\phi$. 

\section{ZAMO observer and local sky}
\label{secZAMO}

The observer of the black hole is chosen to be a zero-angular-momentum observer (ZAMO),
also known as a locally nonrotating observer~\cite{BardeenPressTeukolsky1972}.
This choice is natural.
First, the ZAMO is locally defined and does not require an
asymptotically flat region.  This is essential for a spacetime with a
cosmological horizon or a running cosmological sector.
And, second, its local tetrad provides an unambiguous physical
orthonormal screen.

The observer is placed at
$(r_{\rm O},\theta_{\rm O})$
inside the external type-I block.  In particular,
\begin{equation}
\Deltar(r_{\rm O})>0,
\qquad
\Deltat(r_{\rm O},\theta_{\rm O})>0,
\qquad
F(r_{\rm O},\theta_{\rm O})>0.
\label{eq:ZAMOconditions}
\end{equation}

The ZAMO angular velocity is \cite{Torres2026}
\begin{equation}
\Om
=
-\frac{g_{t\phi}}{g_{\phi\phi}}
=
\frac{
a\left[\Deltat(r^2+a^2)-\Deltar\right]
}{F}.
\label{eq:Omega}
\end{equation}
We introduce the positive functions
\begin{equation}
N^2=\frac{\Delta_r\Delta_\theta\Sigma}{\Xi^2F},
\qquad
B_r^2=\frac{\Sigma}{\Delta_r},
\qquad
B_\theta^2=\frac{\Sigma}{\Delta_\theta},
\qquad
B_\phi^2=\frac{F\sin^2\theta}{\Xi^2\Sigma},
\label{eq:ZAMOfactors}
\end{equation}
An orthonormal ZAMO tetrad is
\begin{equation}
e_{(\hat t)}=N^{-1}\left(\partial_t+\Omega\partial_\phi\right),
\qquad
e_{(\hat r)}=B_r^{-1}\partial_r,
\qquad
e_{(\hat\theta)}=B_\theta^{-1}\partial_\theta,
\qquad
e_{(\hat\phi)}=B_\phi^{-1}\partial_\phi .
\label{eq:ZAMOtetrad}
\end{equation}

Backward ray tracing is performed from the observer toward the black
hole.  We normalize the photon energy measured in the local tetrad to
unity and write
\begin{align}
p^{(\hat t)}&=1,
\label{eq:localpt}\\
p^{(\hat r)}
&=
-\sqrt{1-u^2-v^2},
\label{eq:localpr}\\
p^{(\hat\theta)}&=v,
\label{eq:localpth}\\
p^{(\hat\phi)}&=-u .
\label{eq:localpphi}
\end{align}
From (\ref{eq:localpr}), the variables $(u,v)$ satisfy
They are bounded \emph{direction cosines}, not Bardeen impact
coordinates.  

Let
\begin{equation}
w\equiv\sqrt{1-u^2-v^2}.
\label{eq:w}
\end{equation}
Because the ZAMO tetrad is orthonormal and the locally measured
photon energy has been normalized to $p^{(\hat t)}=1$, the null
condition implies
\begin{equation}
\left(p^{(\hat r)}\right)^2
+
\left(p^{(\hat\theta)}\right)^2
+
\left(p^{(\hat\phi)}\right)^2
=1.
\end{equation}
Thus the spatial photon momentum defines a unit direction in the
observer's local rest space,
\begin{equation}
\boldsymbol n
=
-w\,\boldsymbol e_{\hat r}
+v\,\boldsymbol e_{\hat\theta}
-u\,\boldsymbol e_{\hat\phi}.
\label{eq:localdirection}
\end{equation}
If $\vartheta_{\rm sky}$ denotes the angle between this direction and
the inward radial direction $-\boldsymbol e_{\hat r}$, then
\begin{equation}
\cos\vartheta_{\rm sky}=w,
\qquad
\sin\vartheta_{\rm sky}=\sqrt{u^2+v^2}.
\label{eq:skyangle}
\end{equation}
Consequently, the radial coordinate
$\rho=\sqrt{u^2+v^2}$ on the direction-cosine screen satisfies
$\rho=\sin\vartheta_{\rm sky}$.

Resolving the locally measured photon momentum in the ZAMO tetrad
and converting it to Boyer--Lindquist canonical momenta gives, at the
observer,
\begin{align}
E
&=
N_{\rm O}
-u\,\Om_{\rm O}B_{\phi{\rm O}},
\label{eq:Einitial}\\
L
&=
-u B_{\phi{\rm O}},
\label{eq:Linitial}\\
p_r
&=
-B_{r{\rm O}}\,w,
\label{eq:prinitial}\\
p_\theta
&=
B_{\theta{\rm O}}\,v .
\label{eq:pthetainitial}
\end{align}
Equations~\eqref{eq:Einitial}--\eqref{eq:pthetainitial} 
satisfy the null constraint \eqref{eq:Kzero} identically.

A useful gnomonic, or tangent-plane, representation of the same local
sky is
\begin{equation}
X=\frac{u}{w},
\qquad
Y=\frac{v}{w}.
\label{eq:gnomonic}
\end{equation}
The local direction-cosine plane $(u,v)$ and the gnomonic plane
$(X,Y)$ describe precisely the same photon directions; they merely use
different projections of the celestial sphere.

The finite-distance shadow observables reported below are evaluated
on the local gnomonic screen $(X,Y)$. For comparison with standard
Kerr benchmarks, the mass-normalized asymptotic celestial coordinates
are obtained as
\begin{equation}
(\alpha_B,\beta_B)
=
\lim_{r_O/M\rightarrow\infty}
\frac{r_O}{M}(X,Y),
\label{eq:asymptotic_screen}
\end{equation}
where the limit is taken in the Kerr geometry,
$\ell=\Lambda_0=0$, at fixed observer inclination.
(This relation supplies the normalization for the asymptotic
shadow-size comparisons discussed in Appendix~\ref{app:numericschecks}).

It is interesting to note that, for an equatorial observer ($\theta_{\rm O}=\pi/2$), rays
launched along the horizontal screen axis ($v=0$) have
$p_\theta=0$ and remain in the invariant equatorial plane.
For the configurations studied here, the two unstable equatorial
light rings therefore determine the intersections of the shadow
boundary with that axis. For an inclined observer, a horizontal
screen direction still has $p_\theta=0$ initially, but the ray
generally develops polar motion, so this direct identification
no longer applies.

\section{Shadow construction}
\label{sec:shadowconstruction}

The black-hole shadow consists of directions on the observer's
celestial sphere whose backward-traced null geodesics reach the
black-hole horizon rather than the external illuminating region.
We initialize rays using
Eqs.~\eqref{eq:localpt}--\eqref{eq:pthetainitial} and integrate
the full null Hamilton system
\eqref{eq:rdot}--\eqref{eq:pthetadot} from the observer.

Numerically, capture is identified through near-horizon stopping
criteria, while escape requires reaching an outer source surface
$r=r_{\rm src}$ with outward radial momentum.
For the asymptotically Kerr--de Sitter configurations, this surface
satisfies $r_{\rm O}<r_{\rm src}<r_C$.
Unresolved rays are reintegrated over a longer Mino-parameter
interval with the same step size.

For each sampled screen angle, we bracket and bisect the
capture--escape transition to reconstruct the shadow boundary.
The stopping criteria, treatment of persistently unresolved
rays, and parallel boundary-reconstruction procedure are
specified in Appendix~\ref{app:numerics}.

\section{Shadow observables}
\label{secObservables}

We characterize the shadow boundary $\partial\mathcal{S}$ in the
local gnomonic coordinates $(X,Y)$ introduced above through
observables measuring its size, displacement, and shape.
Let $R=(X_R,Y_R)$, $L=(X_L,Y_L)$, $T=(X_T,Y_T)$ and
$B=(X_B,Y_B)$ denote, respectively, the rightmost, leftmost,
uppermost and lowermost points of the shadow boundary
$\partial\mathcal S$.

For the quantitative characterization of the shadow we use observables
that separate its overall size from its morphology.  A convenient
global measure of size is the shadow area \cite{KumarGhosh2020},
\begin{equation}
A_{\rm sh}
=\frac{1}{2}
\left|\oint_{\partial\mathcal{S}}
\left(X\,dY-Y\,dX\right)\right|,
\end{equation}
from which we define the area-equivalent radius,
\begin{equation}
R_A=\sqrt{\frac{A_{\rm sh}}{\pi}}.
\end{equation}
The quantity $A_{\rm sh}$ (or $R_A$) is the principal measure of the
geometrical shadow size, since global size is the shadow property most
directly connected, after calibration with an emission model, with
present horizon-scale interferometric constraints
\cite{EHTM87VI2019,EHTSgrAVI2022}.  The morphology is further
characterized by the horizontal and vertical diameters
\begin{equation}
W=X_R-X_L,
\qquad
H=Y_T-Y_B,
\end{equation}
the aspect ratio
\begin{equation}
{\cal A}=\frac{H}{W},
\label{eq:aspectratio}
\end{equation}
and the horizontal displacement
\begin{equation}
D_{\rm sh}=\frac{X_R+X_L}{2},
\end{equation}
whose sign depends on the orientation chosen for the horizontal screen
coordinate.  
For comparison with the extensive earlier literature on Kerr shadows, and
as a useful validation of the numerical calculation, we also quote
when appropriate the Hioki--Maeda observables
\cite{HiokiMaeda2009}: 
If $R_s$ denotes the radius of the reference circle passing
through the topmost, bottommost, and rightmost shadow points,
and $X_c$ is the horizontal coordinate of its center, then
$X_{\mathrm{circ},L}=X_c-R_s$ is its leftmost horizontal
coordinate. The dent and distortion parameter are, respectively,
\begin{equation}
D_{\rm cs}=X_L-X_{\mathrm{circ},L},
\qquad
\delta_s=\frac{D_{\rm cs}}{R_s}.
\end{equation}
%

\section{Results}
\label{secResults}

For the numerical analysis it is convenient to express all lengths in
units of the asymptotic black-hole mass $M$.   We introduce the dimensionless
\textit{radial} coordinate $x\equiv r/M$ and the parameters
\begin{equation}
a_\ast=\frac{a}{M},
\qquad
\epsilon=\frac{\ell}{M},
\qquad
\lambda=\Lambda_0 M^2,
\qquad
x_{\rm O}=\frac{r_{\rm O}}{M}.
\label{eq:dimensionlessParameters}
\end{equation}
Unless stated otherwise, the numerical results below are expressed in
terms of $(a_\ast,\epsilon,\lambda,x_{\rm O})$ and the observer
inclination $\theta_{\rm O}$.

Before discussing the running-$\Lambda$ geometries, we validated the
ray-tracing and shadow-extraction procedure against analytically known
limits.  In particular, the Schwarzschild shadow radius
$3\sqrt{3}\,M$ and the Kerr critical curve are recovered to the
numerical accuracy of the calculation.  
Details of these
validation tests and convergence checks are given in
Appendix \ref{app:numerics}.

\subsection{Weak-core-correction limit}

Let us consider here the case in which $\epsilon=\ell/M\ll1$ at fixed $a_*$, $\lambda$, and
observer configuration. Writing $x\equiv r/M$, the dimensionless
profiles $\mu(x)\equiv\mathcal{M}(Mx)/M$ and
$\widehat{\Lambda}(x)\equiv M^2\Lambda(Mx)$ admit the expansions
\begin{align}
\mu(x)
&=1-\frac{\epsilon^3}{x^3}
  +\mathcal{O}\!\left(\frac{\epsilon^6}{x^6}\right),
\\
\widehat{\Lambda}(x)
&=\lambda+\frac{\epsilon^2}{x^4}
  -\lambda\frac{\epsilon^4}{x^4}
  +\mathcal{O}\!\left(\frac{\epsilon^6}{x^8}\right),
\end{align}
as $\epsilon\to0$ at fixed $x>0$. At photon-region radii
$x=\mathcal{O}(1)$, the mass-profile correction therefore
begins at order $\epsilon^3$, while the running cosmological
term contributes already at order $\epsilon^2$.

For the dimensionless radial metric function
$\widehat{\Delta}_r(x)\equiv\Delta_r(Mx)/M^2$, these expansions give
\begin{align}
\widehat{\Delta}_r
={}&x^2+a_*^2-2x
-\frac{\lambda}{3}x^2(x^2+a_*^2)
\notag\\
&-\frac{\epsilon^2}{3}\frac{x^2+a_*^2}{x^2}
+\frac{2\epsilon^3}{x^2}
+\mathcal{O}(\epsilon^4).
\end{align}
The first line is the Kerr--de Sitter expression with the
same asymptotic parameters. The remaining metric functions
are controlled by the same expansion. In particular,
defining
$\delta\widehat{\Lambda}\equiv\widehat{\Lambda}-\lambda$,
we have the exact relations
\begin{equation}
\Delta_\theta-\Delta_\theta^{(0)}
=\frac{a_*^2\cos^2\theta}{3}\,
 \delta\widehat{\Lambda},
\qquad
\Xi-\Xi^{(0)}
=\frac{a_*^2}{3}\,\delta\widehat{\Lambda},
\end{equation}
where the superscript $(0)$ denotes the $\epsilon=0$ limit,
while $\Sigma$ is independent of $\epsilon$.
Thus, $\Delta_\theta$ and $\Xi$ also receive corrections
of order $\epsilon^2$.

Since $F$, the inverse metric, and the observer's tetrad
follow algebraically from these functions, the same hierarchy
extends to the full null Hamiltonian and the local screen
construction on compact exterior regions where these
quantities remain regular. It is also preserved by the
derivatives entering Hamilton's equations. 
For a fixed Kerr--de Sitter reference configuration
($\epsilon=0$) away from degeneracies of the trapped photon dynamics,
the critical curve responds smoothly to these perturbations, and the
shadow observables consequently inherit perturbatively small corrections.

Thus, a regularization length comparable, for example, to the Planck length would produce strongly suppressed deviations from the Kerr–de Sitter shadow for black holes of stellar mass or larger.
Appreciable departures require an effective value of $\ell/M$ that is
not negligibly small.
 At present, however, observations do not uniquely
determine such a strong-field regularization scale. In particular,
constraints on the low-curvature evolution of the dark-energy sector
do not, by themselves, determine its value. Moreover,
different quantum-gravity-inspired constructions can associate the onset of
effective modifications with substantially different length scales~\cite{RovelliVidotto2014,Mathur2006,Giddings2014}.

\subsection{Separate effects of the running mass and cosmological terms}
\label{subsec:running_M_Lambda_shadow}

It is instructive to disentangle the respective roles of the radial
dependences of the mass and cosmological terms. To this end, we compare
three configurations with the same spin, asymptotic cosmological
constant, and observer position,
$a_*=0.8$, $\lambda=\Lambda_0 M^2=10^{-6}$, $x_{\rm O}=30$,
and the same observer inclination and screen prescription.  The three
cases are: (i) classical Kerr--de Sitter (KdS) solution, corresponding to
$\epsilon=0$; (ii) a constant mass $M$ with the running cosmological
term
\begin{equation}
 \widehat{\Lambda}(x)
 =
 \frac{\epsilon^2+\lambda x^4}{\epsilon^4+x^4},
 \qquad \epsilon=0.83;
 \label{eq:runningLambda_comparison}
\end{equation}
and (iii) the genuine non-singular rotating black hole model that has the same running cosmological term together with a running effective mass
\begin{equation}
\mathcal M(x)=M\mu(x),
 \qquad
 \mu(x)=\frac{x^3}{x^3+\epsilon^3}.
 \label{eq:runningM_comparison}
\end{equation}
The resulting horizon, light-ring, and shadow observables are summarized
in Table~\ref{tab:running_M_Lambda_comparison}.

\begin{table*}[t]
\centering
\caption{Comparison of the Kerr--de Sitter geometry (KdS), the geometry
with running $\Lambda(r)$ and constant $\mathcal M$ ($\Lambda$-run), and the
geometry in which both $\mathcal M(r)$ and $\Lambda(r)$ run ($M+\Lambda$-run).
For an observable ${\cal O}$, the percentages are defined by
$\Delta_\Lambda=100({\cal O}_{\Lambda}/{\cal O}_{\rm KdS}-1)$,
$\Delta_{\rm tot}=100({\cal O}_{M+\Lambda}/{\cal O}_{\rm KdS}-1)$,
and
$\Delta_M=100({\cal O}_{M+\Lambda}/{\cal O}_{\Lambda}-1)$.
Thus, $\Delta_M$ isolates the additional effect of introducing the
running mass while keeping the same running cosmological term.
For the retrograde impact parameter the percentage variations refer
to its absolute value.}
\label{tab:running_M_Lambda_comparison}
\begin{tabular}{lrrrrrr}
\hline\hline
Observable
& KdS
& $\Lambda$-run
& $\Delta_\Lambda$ [\%]
& $M+\Lambda$-run
& $\Delta_{\rm tot}$ [\%]
& $\Delta_M$ [\%]
\\
\hline
$x_+$
& 1.60000 & 1.78952 & $+11.84$
& 1.34897 & $-15.69$ & $-24.62$
\\
$x_{\rm C}$
& 1731.04994 & 1731.04987 & $\simeq 0$
& 1731.04987 & $\simeq 0$ & $\simeq 0$
\\[1mm]

$W$
& 0.32424 & 0.34187 & $+5.44$
& 0.31887 & $-1.66$ & $-6.73$
\\
$H$
& 0.33984 & 0.35373 & $+4.09$
& 0.34735 & $+2.21$ & $-1.80$
\\
$D_{\rm sh}$
& 0.05733 & 0.05404 & $-5.75$
& 0.06384 & $+11.35$ & $+18.15$
\\
${\cal A}$
& 1.04811 & 1.03467 & $-1.28$
& 1.08929 & $+3.93$ & $+5.28$
\\
$R_s$
& 0.16997 & 0.17690 & $+4.08$
& 0.17374 & $+2.21$ & $-1.79$
\\
$D_{cs}$
& 0.01570 & 0.01192 & $-24.07$
& 0.02860 & $+82.15$ & $+139.88$
\\
$\delta_s$
& 0.09237 & 0.06739 & $-27.04$
& 0.16460 & $+78.20$ & $+144.25$
\\
$A_{\rm sh}$
& 0.08684 & 0.09517 & $+9.59$
& 0.08865 & $+2.08$ & $-6.85$
\\
$R_A$
& 0.16626 & 0.17405 & $+4.68$
& 0.16798 & $+1.04$ & $-3.48$
\\[1mm]

$r_{\rm LR}^{-}/M$
& 1.81109 & 2.07504 & $+14.57$
& 1.41695 & $-21.76$ & $-31.71$
\\
$b_{\rm LR}^{-}$
& 3.23730 & 3.60720 & $+11.43$
& 2.95646 & $-8.68$ & $-18.04$
\\
$r_{\rm LR}^{+}/M$
& 3.81876 & 3.95243 & $+3.50$
& 3.88915 & $+1.84$ & $-1.60$
\\
$|b_{\rm LR}^{+}|$
& 6.66255 & 6.82316 & $+2.41$
& 6.77416 & $+1.68$ & $-0.72$
\\
\hline\hline
\end{tabular}
\end{table*}

\begin{figure}[ht]
    \centering
    \includegraphics[scale=0.8]
        {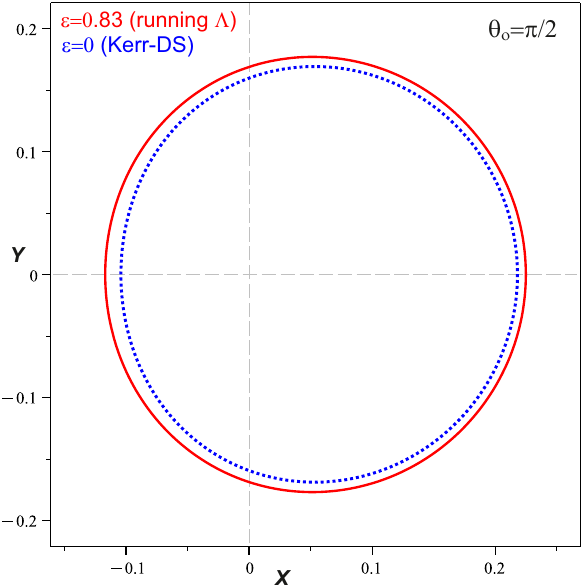}
    \caption{
    Comparison between the shadow of the Kerr--de Sitter reference
    geometry ($\epsilon=0$, blue dashed curve) and that obtained when
    only the cosmological term is allowed to run
    ($\epsilon=0.83$, red solid curve), while the mass parameter is
    kept constant.  In both cases
    $a_*=0.8$, $\lambda=\Lambda_0M^2=10^{-6}$,
    $x_{\rm O}=30$, and $\theta_{\rm O}=\pi/2$.
    The running cosmological term increases the overall angular size
    of the shadow while reducing its spin-induced deformation,
    particularly on the prograde side.  Quantitatively, this behavior
    corresponds to an increase in the characteristic shadow area and vertical diameter
    together with a decrease in the distortion parameter
    $\delta_s$, as summarized in
    Table~\ref{tab:running_M_Lambda_comparison}.
    }
    \label{fig:shadow_KdS_runningLambda}
\end{figure}

\begin{figure}[ht]
    \centering
    \includegraphics[scale=0.9]{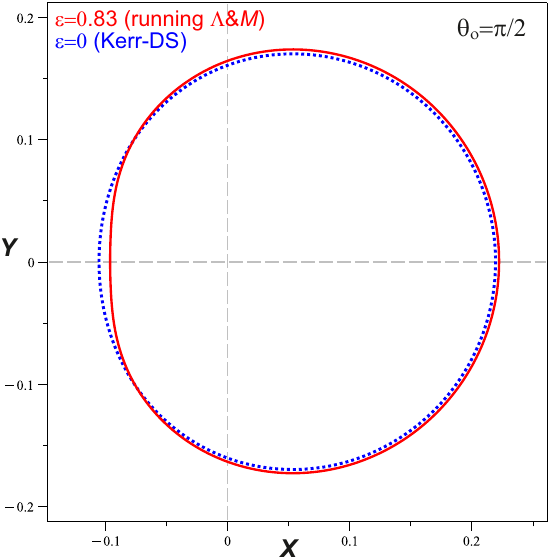}
    \caption{
    Shadow contours for the Kerr--de Sitter reference geometry
    ($\epsilon=0$, blue dashed curve) and for the geometry with both
    $\mathcal M(r)$ and $\Lambda(r)$ running
    ($\epsilon=0.83$, red solid curve).
    In both cases
    $a_*=0.8$, $\lambda=\Lambda_0M^2=10^{-6}$,
    $x_{\rm O}=30$, and $\theta_{\rm O}=\pi/2$.
    With the chosen parameters, the running geometry is nearly extremal and produces a markedly stronger spin-induced deformation, most visible on the prograde side, while changing
    the overall characteristic shadow size.
    The differential modification reflects the stronger influence of
    the running functions on the innermost prograde photon sector.
    Quantitative values are reported in
    Table~\ref{tab:running_M_Lambda_comparison}.
    }
    \label{fig:shadow_KdS_runningML}
\end{figure}

The comparison exhibits a clear competition between the two running
terms.  When only $\Lambda(r)$ is allowed to run, the shadow becomes
larger while its rotational distortion decreases.  
In particular, the shadow area $A_{\rm sh}$ (equivalently,
the area-equivalent radius $R_A$) and the Hioki--Maeda
reference-circle radius $R_s$ increase, whereas the dent
$D_{\rm cs}$ and the distortion parameter $\delta_s$ decrease. (See Table~\ref{tab:running_M_Lambda_comparison} and Figure~\ref{fig:shadow_KdS_runningLambda}). This behavior can be
traced to the strongly radius-dependent character of
Eq.~\eqref{eq:runningLambda_comparison}.  Although
$\lambda$ is small asymptotically,
$\widehat{\Lambda}(x)$ can be much larger in the strong-field region\footnote{This explains why the qualitative effects found on the shadow persist for accurate astrophysically small values of the asymptotic cosmological constant.}.
The correction therefore acts much more strongly on the prograde
critical photon trajectories, which probe smaller radii, than on
their retrograde counterparts.
This differential response is apparent already from the equatorial
light rings.  Running $\Lambda(r)$ moves the prograde light ring
outwards considerably more than the retrograde one.  It therefore
reduces the radial contrast between the two branches of the trapped
photon dynamics.  Since the characteristic Kerr deformation of the
shadow originates from the different behavior of the co-rotating and
counter-rotating photon sectors, the preferential outward displacement
of the former partly compensates the asymmetry induced by frame
dragging.  
We, thus, conclude that the running cosmological term tends to
make the shadow larger but less asymmetric.

The effect of the running mass is qualitatively different.  Keeping the running $\Lambda(r)$ and replacing $M$ by Eq.~\eqref{eq:runningM_comparison}
produces a pronounced inward displacement of the event horizon and,
even more significantly, of the prograde light ring, whereas the
retrograde light ring changes comparatively little.  (See Table~\ref{tab:running_M_Lambda_comparison} and Figure~\ref{fig:shadow_KdS_runningML}).  This behavior
follows directly from the radial localization of the mass
regularization.  Since $\mu(x)<1$ and $\mu(x)\longrightarrow 1$
for $x\gg\epsilon$,
the effective gravitational mass differs substantially from its
asymptotic value only in the innermost region.  
For a regularization scale comparable to the black-hole mass,
$\ell/M={\cal O}(1)$, the prograde photon sector therefore probes a
significantly reduced effective mass, whereas the retrograde sector,
located farther out, already samples a geometry for which
$\mathcal{M}(r)\simeq M$.
The running mass consequently
enhances, rather than compensates, the difference between the two
sides of the photon region.
The competition can also be seen directly in the radial metric
function.  Relative to a constant-mass geometry, the mass profile
produces
\begin{equation}
 \delta\widehat{\Delta}_{M}
 =
 2x\,[1-\mu(x)]>0 ,
 \label{eq:deltaDeltaM}
\end{equation}
whereas, relative to KdS, the enhancement of the cosmological term in
the strong-field region gives
\begin{equation}
 \delta\widehat{\Delta}_{\Lambda}
 =
 -\frac{\widehat{\Lambda}(x)-\lambda}{3}
 x^2(x^2+a_*^2)<0 .
 \label{eq:deltaDeltaLambda}
\end{equation}
The two contributions thus enter $\widehat{\Delta}_r$ with opposite
signs.  The photon dynamics depends also on their radial derivatives,
so Eqs.~\eqref{eq:deltaDeltaM} and
\eqref{eq:deltaDeltaLambda} should not by themselves be interpreted
as determining the displacement of the critical orbits.  They do,
however, make explicit the competing character of the two
modifications observed in the numerical results.

The consequence for the shadow is particularly clear.  Adding the
running mass to the running-$\Lambda$ geometry partly reverses the
increase in the overall shadow size, but it more than reverses the
reduction in distortion.  The distortion parameter $\delta_s$ becomes
substantially larger than not only its value in the
running-$\Lambda$ configuration but also its KdS value.  The same
conclusion follows from the absolute dent $D_{cs}$ and from the
horizontal displacement $D_{\rm sh}$, showing that the effect is not
merely a consequence of the normalization entering the definition of
$\delta_s$.  Thus, for the parameters considered here, the running
cosmological term and the running mass have opposite effects on the
spin-induced deformation of the shadow: the former partially
suppresses it, whereas the latter strongly enhances it.
Nevertheless, let us clarify that the combined modification is intrinsically nonlinear. 
The final shadow cannot consequently be understood as a simple linear
superposition of independent ``mass-running'' and
``cosmological-running'' corrections.

Finally, the cosmological horizon is practically identical in all
three configurations.  This provides a useful consistency check and
illustrates the separation of scales built into the model.  At large
radius, $\mathcal M(r)\longrightarrow M$, $\Lambda(r)\longrightarrow\Lambda_0$,
so that the geometries approach the same asymptotic Kerr--de Sitter
solution, while their horizon-scale and photon-region properties can
remain markedly different.  The comparison therefore shows explicitly
that sizeable changes in black-hole optical observables can coexist
with an essentially unchanged asymptotic Kerr-de Sitter sector.

\subsection{Dependence on the regularization scale and spin}
\label{subsec:parameter_scans}

Table~\ref{tab:parameter_scans} extends the fiducial comparison through
separate scans in $\epsilon$ and $a_*$, at fixed
$\lambda$, $x_{\rm O}$, and $\theta_{\rm O}$.
Both $\mathcal M(r)$ and $\Lambda(r)$ run for $\epsilon>0$.

\begin{table*}[t]
\centering
\caption{Shadow observables on the local gnomonic screen for
$\lambda=10^{-6}$, $x_{\rm O}=30$, and $\theta_{\rm O}=\pi/2$.
The upper block varies $\epsilon$ at $a_*=0.8$ and includes the
Kerr--de Sitter reference ($\epsilon=0$); the lower block varies
$a_*$ at $\epsilon=0.83$. Both profiles run when $\epsilon>0$.}
\label{tab:parameter_scans}
\small
\setlength{\tabcolsep}{3.5pt}
\begin{tabular}{@{}lrrrrrrrrr@{}}
\toprule
 & $W$ & $H$ & $D_{\rm sh}$ & ${\cal A}$ & $R_s$
 & $D_{cs}$ & $\delta_s$ & $A_{\rm sh}$ & $R_A$ \\
\midrule
\multicolumn{10}{@{}l}{Regularization-scale scan: $a_*=0.8$;
first column gives $\epsilon$} \\
\addlinespace[2pt]
$0$ (KdS) & 0.32424 & 0.33984 & 0.05733 & 1.04811 & 0.16997 & 0.01570 & 0.09237 & 0.08684 & 0.16626 \\
$0.41$ & 0.32679 & 0.34253 & 0.05722 & 1.04817 & 0.17132 & 0.01584 & 0.09247 & 0.08824 & 0.16759 \\
$0.62$ & 0.32718 & 0.34499 & 0.05825 & 1.05442 & 0.17255 & 0.01791 & 0.10380 & 0.08913 & 0.16843 \\
$0.83$ & 0.31887 & 0.34735 & 0.06384 & 1.08929 & 0.17374 & 0.02860 & 0.16460 & 0.08865 & 0.16798 \\
\midrule
\multicolumn{10}{@{}l}{Spin scan: $\epsilon=0.83$;
first column gives $a_*$} \\
\addlinespace[2pt]
$0.4$ & 0.34192 & 0.34597 & 0.02796 & 1.01185 & 0.17299 & 0.00406 & 0.02349 & 0.09293 & 0.17199 \\
$0.6$ & 0.33622 & 0.34654 & 0.04316 & 1.03072 & 0.17330 & 0.01038 & 0.05988 & 0.09169 & 0.17084 \\
$0.8$ & 0.31887 & 0.34735 & 0.06384 & 1.08929 & 0.17374 & 0.02860 & 0.16460 & 0.08865 & 0.16798 \\
\bottomrule
\end{tabular}
\end{table*}

Both the area and vertical diameter exceed their Kerr--de Sitter
values at the sampled nonzero values of $\epsilon$.
Their responses nevertheless differ: $H$ increases across
the scan, whereas $A_{\rm sh}$ is largest at
$\epsilon=0.62$ among the sampled configurations.
On the other hand, the width exceeds its Kerr--de Sitter value at
$\epsilon=0.41$ and $0.62$, but falls below it at $0.83$;
the distortion is nearly unchanged at $0.41$ and strongly enhanced
at $0.83$.

At fixed $\epsilon=0.83$, increasing $a_*$ from $0.4$ to $0.8$
reduces the area by $4.61\%$ while increasing the distortion by
a factor of approximately seven. The vertical diameter changes
by only $0.40\%$. Thus, the pronounced deformation of the
nearly extremal fiducial configuration becomes substantially
weaker at the lower spins considered here.

\subsection{Dependence on observer inclination}
\label{subsec:inclination_dependence}

Table~\ref{tab:inclination_running_shadow} varies the inclination
within the same running geometry. Lowering $\theta_{\rm O}$ from
$\pi/2$ to $0.2\,{\rm rad}$ reduces the area by only $2.09\%$,
whereas the displacement and distortion decrease by factors of
approximately $5.3$ and $24.5$, respectively.
The intermediate inclination follows these trends, with the
aspect ratio approaching unity towards the polar view.

\begin{table*}[t]
\centering
\caption{Inclination dependence for the geometry with both
$\mathcal M(r)$ and $\Lambda(r)$ running, at
$a_*=0.8$, $\epsilon=0.83$, $\lambda=10^{-6}$, and $x_{\rm O}=30$.
Angles are in radians. The intrinsic horizon and equatorial
light-ring quantities are independent of inclination and are
given in Table~\ref{tab:running_M_Lambda_comparison}.}
\label{tab:inclination_running_shadow}
\small
\setlength{\tabcolsep}{3.5pt}
\begin{tabular}{@{}lrrrrrrrrr@{}}
\toprule
$\theta_{\rm O}$ & $W$ & $H$ & $D_{\rm sh}$ & ${\cal A}$ & $R_s$
& $D_{cs}$ & $\delta_s$ & $A_{\rm sh}$ & $R_A$ \\
\midrule
$\pi/2$ & 0.31887 & 0.34735 & 0.06384 & 1.08929 & 0.17374 & 0.02860 & 0.16460 & 0.08865 & 0.16798 \\
$\pi/4$ & 0.32630 & 0.34043 & 0.04395 & 1.04331 & 0.17025 & 0.01420 & 0.08338 & 0.08769 & 0.16707 \\
$0.2$ & 0.33188 & 0.33300 & 0.01207 & 1.00336 & 0.16650 & 0.00112 & 0.00671 & 0.08680 & 0.16622 \\
\bottomrule
\end{tabular}
\end{table*}

\begin{figure}[ht]
\centering
\includegraphics[scale=0.7]{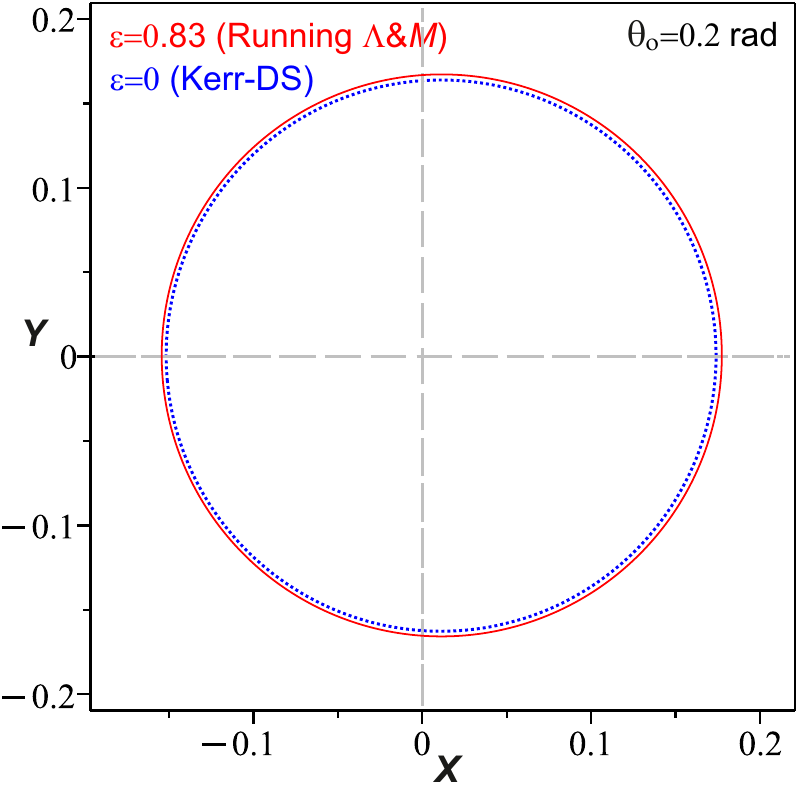}
\caption{Nearly polar shadow contours at $\theta_{\rm O}=0.2\,{\rm rad}$
for Kerr--de Sitter ($\epsilon=0$, blue dotted curve) and the full
running geometry ($\epsilon=0.83$, red solid curve).
Both use $a_*=0.8$, $\lambda=10^{-6}$, $x_{\rm O}=30$, and the
same local observer and screen prescription.}
\label{fig:shadow_theta02_KdS_running}
\end{figure}

Figure~\ref{fig:shadow_theta02_KdS_running} compares the nearly polar
shadow with Kerr--de Sitter at the same asymptotic parameters and
observer configuration. The running geometry increases
$R_s$ and $R_A$ by approximately $2.2\%$ and the area by $4.4\%$,
while the distortion remains small in absolute terms:
$\delta_s\simeq0.00671$, compared with $0.00488$ in Kerr--de Sitter.
The size difference therefore persists even when the absolute
morphological differences are suppressed relative to the
equatorial comparison in Figure~\ref{fig:shadow_KdS_runningML}.

\smallskip
Taken together, these results motivate the following remark.

\emph{Remark.} The simultaneous increase in shadow area and
vertical diameter relative to Kerr--de Sitter is driven primarily
by the running cosmological term and persists across the sampled
regularization scales, spins and observer inclinations.

\section{Conclusions}
\label{secConclu}

We have developed a formulation of photon trapping and black-hole
shadows that applies directly to nonsingular rotating black  holes embedded in
dark-energy geometries \cite{Torres2026}.

The null-geodesic problem has been
written as the four-dimensional first-order Hamiltonian system
\eqref{eq:rdot}--\eqref{eq:pthetadot}, with a Mino-type evolution
parameter.  
Clearly, this formulation requires only stationarity and axial
symmetry and therefore remains valid when the Hamilton--Jacobi
equation is not separable.

For a constant cosmological term, the null equations separate
and admit a Carter-type constant. In the subextremal Kerr
and Kerr--de Sitter limits, the exterior trapped set consists
of spherical photon orbits.

For a running cosmological term, the radial and polar dynamics
are generically coupled. Under the stated exterior regularity
and positivity assumptions, we prove that, for $a\neq0$,
any spherical null geodesic at a radius where
$\Lambda'(r)\neq0$ must be an equatorial light ring.
A numerically resolved orbit, periodic in the reduced phase
space and exhibiting both radial and polar motion, provides
an explicit example of nonspherical trapping.

Consequently, the full shadow boundary cannot be reconstructed
from spherical photon orbits alone. We instead construct the
shadow by direct integration of the full null Hamilton equations
and numerical localization of the capture--escape transition
on the observer's sky.
The equatorial light rings are computed separately from
the circular-orbit conditions, with their radial stability
determined by the equatorial radial potential.
Their radii and impact parameters provide complementary
diagnostics of the running profiles.

The observer has been chosen to be a ZAMO in the external type-I
block.  This local prescription is particularly appropriate for a
spacetime with a nonzero or running cosmological term, since it does
not rely on the existence of asymptotic flatness.  The resulting
direction-cosine screen is physically local and bounded.

We have shown that potentially observable effects require an effective regularization
scale that is not negligibly small compared with $M$.
Our calculations quantify how the effective regularization scale
$\ell$ could affect photon trapping and shadow observables.
Proposed descriptions of nonsingular interiors
and quantum black-hole structure involve different characteristic
lengths, including scales that depend on the black-hole
mass~\cite{RovelliVidotto2014,Mathur2006}.
Cosmological constraints from DESI probe the low-curvature
expansion history and do not, by themselves, determine $\ell$
in the present geometry. Horizon-scale observations offer a
more direct test: analyses of asymptotically flat rotating
regular metrics find that regularization parameters corresponding
to sizeable fractions of $M$ can satisfy the 2022 EHT shadow
constraints on Sgr~A*~\cite{WaliaGhoshMaharaj2022}.
Our calculations provide the shadow predictions needed to
investigate such constraints in a geometry with a running
cosmological term.

Comparisons with Kerr--de Sitter at fixed asymptotic parameters
and observer configuration quantify the effects of the running
mass and cosmological profiles on the shadow observables and
the equatorial light-ring radii and impact parameters.
Specifically, we found
a simultaneous increase in shadow area and vertical diameter,
driven primarily by the running cosmological term and persisting
across the regularization scales, spins and observer inclinations examined.
From an observational perspective, $A_{\rm sh}$ and $H$ are
promising diagnostics of overall shadow size, which may be
more accessible to image reconstruction and interferometric
model fitting than small contour distortions.
Their interpretation requires the relation between the geometric
shadow and the observed emission to be appropriately
modelled~\cite{EHTM87VI2019,EHTSgrAVI2022}.

The modifications introduced by the present
geometry could affect simultaneously the photon
region, the shadow, the ISCO and other observables associated with null and
timelike geodesics.  This opens a realistic possibility of constraining, and
potentially measuring, the effective regularization scale $\ell$ through a joint analysis of complementary
strong-field observables rather than through the shadow alone.  The different
dependence of these observables on $M$, $a$ and $\ell$ is particularly valuable,
since it provides a route to breaking degeneracies that may be present in any
single observable.
The prospects become substantially stronger when several black holes are
considered simultaneously.  A multi-source analysis, spanning objects with
different masses, spins and viewing geometries, would require the same
underlying prescription for the regularization scale to account consistently
for all the observations.  
One may test, for example, whether the regularization scale $\ell$ obeys
a universal mass-dependent scaling.
Such correlated predictions across distinct sources provide a considerably more discriminating
test than fitting an independent deformation parameter to each object.
Consequently, joint analyses of complementary observables
across multiple black holes offer a promising route to placing
stringent bounds on the regularization scale, testing the
strong-field running of the cosmological term, and,
with sufficient observational precision and control of parameter
degeneracies, statistically distinguishing the present regular
rotating geometry from Kerr or Kerr--de Sitter.

\appendix
\section*{Appendix}
\addcontentsline{toc}{section}{Appendix}

\section{Present-epoch reconstruction of \texorpdfstring{$\Lambda(K)$}{Lambda(K)}}
\label{app:LambdaK}

We reconstruct a local relation between the dark-energy density
and the Kretschmann scalar within a spatially flat FLRW model
containing pressureless matter and separately conserved dark energy.
The purpose is to obtain a phenomenological estimate of the
present-epoch dependence of the effective cosmological term on
curvature.
We express the dark-energy energy density
$\varepsilon_{\rm DE}$ as an effective cosmological term,
$\Lambda(z)\equiv 8\pi G\varepsilon_{\rm DE}(z)/c^4$.

For the Chevallier--Polarski--Linder
parametrization~\cite{ChevallierPolarski2001,Linder2003},
energy conservation gives
\begin{equation}
\begin{aligned}
w(z) &= w_0+w_a\frac{z}{1+z},\\
\Lambda(z) &= \Lambda_0 f(z),\\
f(z) &= (1+z)^{3(1+w_0+w_a)}
        \exp\!\left[-\frac{3w_a z}{1+z}\right],
\end{aligned}
\label{eq:CPLbackground}
\end{equation}
where $\Lambda_0=3H_0^2\Omega_{\rm DE,0}/c^2$ and
$\Omega_{\rm DE,0}=1-\Omega_{m0}$. Defining $E(z)=H(z)/H_0$,
the background expansion and curvature are
\begin{equation}
\begin{aligned}
E^2(z) &=
\Omega_{m0}(1+z)^3+\Omega_{\rm DE,0}f(z),\\
q(z) &=
\frac12\left[
1+\frac{3w(z)\Omega_{\rm DE,0}f(z)}{E^2(z)}
\right],\\
K(z) &= \frac{12H_0^4}{c^4}E^4(z)\bigl[1+q^2(z)\bigr],
\end{aligned}
\label{eq:FLRWcurvature}
\end{equation}
where $q=-1-\dot H/H^2$ is the deceleration parameter and
the dot denotes differentiation with respect to cosmic time.

Provided $K'(0)\neq0$, the redshift can be eliminated locally
around the present epoch. Writing $K_0=K(0)$, one obtains
\begin{equation}
\boxed{\frac{\Lambda(K)}{\Lambda_0}
=
1+\beta\left(\frac{K}{K_0}-1\right)
+\mathcal O\!\left[
\left(\frac{K}{K_0}-1\right)^2
\right]},
\label{eq:LambdaKlocal}
\end{equation}
with
\begin{equation}
\begin{aligned}
\beta
&\equiv
\left.\frac{d\ln\Lambda}{d\ln K}\right|_{z=0}
=
\frac{3(1+w_0)}
{4(1+q_0)+2q_0q'_0/(1+q_0^2)},\\
q_0 &= \frac12\bigl[1+3w_0\Omega_{\rm DE,0}\bigr],\\
q'_0 &= \frac32\Omega_{\rm DE,0}
\bigl(w_a+3\Omega_{m0}w_0^2\bigr).
\end{aligned}
\label{eq:LambdaKslope}
\end{equation}
Here a prime denotes differentiation with respect to redshift.
The slope $\beta$ is independent of $H_0$, which sets the
dimensional normalizations $\Lambda_0$ and $K_0$.

For the DESI DR2 BAO+Ly$\alpha$ full-shape+CMB+DES-Dovekie
combination~\cite{DESILya2026}, the published constraints are
\begin{equation}
\begin{aligned}
\Omega_{m0} &= 0.3134\pm0.0052,\\
H_0 &= (67.37\pm0.54)\,\mathrm{km\,s^{-1}\,Mpc^{-1}},\\
w_0 &= -0.821\pm0.054,\qquad
w_a=-0.65\pm0.20.
\end{aligned}
\label{eq:DESIinputs}
\end{equation}
Their central values give
$K_0\simeq3.78\times10^{-104}\,\mathrm{m}^{-4}$ and
$\Lambda_0\simeq1.09\times10^{-52}\,\mathrm{m}^{-2}$.

To estimate the uncertainty in $\beta$, we approximate the
joint distribution of $(\Omega_{m0},w_0,w_a)$ as Gaussian.
The $w_0$--$w_a$ covariance is reconstructed from the published
pivot constraint $w_p=-0.981\pm0.022$ at
$z_p=0.32$~\cite{DESILya2026}. Since
$w_p=w_0+s_pw_a$, with $s_p=z_p/(1+z_p)$,
\[
\operatorname{Cov}(w_0,w_a)
=
\frac{\sigma_{w_p}^2-\sigma_{w_0}^2
-s_p^2\sigma_{w_a}^2}{2s_p},
\]
which yields $\operatorname{Corr}(w_0,w_a)\simeq-0.913$.
Numerical propagation through Eq.~\eqref{eq:LambdaKslope},
neglecting correlations with $\Omega_{m0}$, gives
\begin{equation}
\beta=0.204^{+0.029}_{-0.035}
\qquad (68\%).
\label{eq:betaDESI}
\end{equation}
The quoted bounds represent an approximate equal-tailed
interval for the slope, rather than an interval derived
from the complete DESI posterior chains.

Equation~\eqref{eq:LambdaKlocal}, with the slope $\beta$ given in
Eq.~\eqref{eq:betaDESI}, characterizes the local present-epoch
dependence of the effective cosmological term on curvature along the
assumed FLRW evolution.

\section{Numerical implementation}
\label{app:numerics}

The calculations were implemented in Maple using the dimensionless
variables introduced in the main text. This appendix summarizes
the numerical integration, boundary extraction and reference
settings of the implementation.

\subsection{Integration and ray classification}

Horizon locations are obtained by scanning
$\widehat{\Delta}_r(x)\equiv\Delta_r(Mx)/M^2$ on a logarithmic
grid and refining sign-changing brackets by bisection.
For the configurations with $\Lambda_0>0$ considered here,
the two outermost identified roots define
$x_+=r_+/M$ and $x_C=r_C/M$. The observer is placed in the
exterior stationary region, $x_+<x_O<x_C$, where the metric
conditions required for the ZAMO tetrad are checked.

The observer's tetrad coefficients are evaluated once.
For each local direction $(u,v)$, the initial momenta are
constructed with unit local photon energy. The resulting
conserved quantities $E$ and $L$ are held fixed while
$(x,\theta,p_x,p_\theta)$ are evolved with a fixed-step,
fourth-order Runge--Kutta scheme in the Mino parameter.
Polar-coordinate crossings are handled by reflecting
$\theta$ into $[0,\pi]$ and reversing $p_\theta$.

At each integration step, capture is assigned when
\begin{equation}
x\leq x_++\delta_H
\quad\hbox{or}\quad
\bigl[\widehat{\Delta}_r(x)\leq\tau_\Delta
\ \hbox{and}\ p_x<0\bigr],
\label{eq:numericalCapture}
\end{equation}
where $\delta_H$ is the dimensionless horizon buffer and
$\tau_\Delta$ is the radial-metric threshold.
Escape is assigned when $x\geq x_{\rm esc}$ and $p_x>0$, with
\begin{equation}
x_{\rm esc}
=
\min\!\left[
2x_O,\,
x_O+\frac{x_C-x_O}{4}
\right].
\label{eq:numericalEscape}
\end{equation}
For $\Lambda_0=0$, the implementation uses $x_{\rm esc}=2x_O$.

If neither stopping condition is met before the initial
Mino-parameter integration limit $\gamma_{\max}$, an additional
near-horizon rule assigns capture when the minimum sampled radius
satisfies $x_{\min}\leq x_+ + 2\delta_H$. Otherwise, the ray is
classified as unresolved. An unresolved ray is reintegrated from its
initial conditions with the integration limit extended to
$2\gamma_{\max}$, keeping the step size unchanged.

\subsection{Boundary reconstruction and observables}

The boundary search assumes a single-valued polar
representation $\rho_{\rm sh}(\psi)$ on the local direction
screen,
\[
(u,v)=\rho(\cos\psi,\sin\psi),
\qquad 0\leq\rho<1.
\]
Its applicability is therefore restricted to shadows that
are star-shaped with respect to the screen origin.
For each angle, the radial scan identifies the first
capture-to-escape transition and constructs a bracket with
classified endpoints.

The bracket is narrowed by bisection until its width is at
most $\epsilon_\rho$. If the midpoint remains unresolved
after the extended integration, the algorithm tests interior
points at one-quarter and, if needed, three-quarters of the
bracket. Endpoints are updated only using classified rays.
A bracket of width at most $4\epsilon_\rho$ is accepted
when its midpoint remains unresolved or the iteration limit
is reached. Its midpoint is then stored as the boundary
estimate. A direction for which no acceptable bracket is
obtained is reported as failed.

Boundary calculations proceed in parallel passes. An
initial set of uniformly spaced seed directions provides
a periodic cubic interpolant, which predicts local search
windows for the full angular grid,
\[
\psi_j=\frac{(2j+1)\pi}{N_\psi},
\qquad j=0,\ldots,N_\psi-1.
\]
Unsuccessful local searches are repeated in wider windows
and, if necessary, over the full radial search interval.
A further pass repeats the boundary calculation using the
previous contour as predictor. Each pass is completed before
the next begins, and a failed direction terminates the
calculation before the final interpolation.

The accepted boundary samples define a periodic $C^2$
cubic spline for $\rho_{\rm sh}(\psi)$. After transformation
to the gnomonic screen, the characteristic extrema are
located by uniform angular sampling followed by
golden-section refinement. The remaining shape observables
follow from these points. The area is evaluated by composite
Simpson quadrature in the form
\begin{equation}
A_{\rm sh}
=
\frac12\int_0^{2\pi}
\frac{\rho_{\rm sh}^2(\psi)}
     {1-\rho_{\rm sh}^2(\psi)}
\,d\psi,
\label{eq:numericalShadowArea}
\end{equation}
and $R_A=\sqrt{A_{\rm sh}/\pi}$.

\subsection{Numerical settings and consistency checks}
\label{app:numericschecks}

Table~\ref{tab:numericalSettings} lists the reference production
settings. With $N_{\rm w}$ worker processes, the seed sample
contains
$\min\{N_\psi,\max(24,2N_{\rm w})\}$ directions.

\begin{table}[htbp]
\centering
\caption{Reference numerical settings of the implementation.
The angular sampling used for post-processing is distinct
from the number of ray-traced boundary directions.}
\label{tab:numericalSettings}
\begin{tabular}{lc}
\hline\hline
Quantity & Value\\
\hline
Boundary directions, $N_\psi$ & $360$\\
Mino-parameter step, $h_\gamma$ & $2.5\times10^{-4}$\\
Initial integration limit, $\gamma_{\max}$ & $50$\\
Boundary tolerance, $\epsilon_\rho$ & $5\times10^{-7}$\\
Horizon buffer, $\delta_H$ & $5\times10^{-3}$\\
Radial-metric threshold, $\tau_\Delta$ & $10^{-7}$\\
Maximum bisection iterations & $60$\\
Additional boundary passes & $1$\\
Logarithmic root-search intervals & $12000$\\
Angular samples for locating extrema & $20000$\\
Simpson integration panels & $20000$\\
\hline\hline
\end{tabular}
\end{table}

The code reports the periodic closure error and the maximum
difference between the final spline and its boundary samples.
These quantities monitor interpolation consistency.

We tested the sensitivity to the Mino-parameter integration step for the representative configuration
$a_\ast=0.8$, $\epsilon=0.83$, $\lambda=10^{-6}$, $x_O=30$, and
$\theta_O=\pi/2$, using
$h_\gamma=5.0\times10^{-4}$, $2.5\times10^{-4}$, and
$1.25\times10^{-4}$, while keeping all other numerical parameters fixed.
For each observable $O$ we define
$\Delta_hO=\max_i O(h_i)-\min_i O(h_i)$ and
$\delta_hO=\Delta_hO/|O(h_{\min})|$.
All observables are extremely stable under this factor-four variation of
$h_\gamma$: the largest relative variations occur for $D_{\rm cs}$ and
$\delta_s$, reaching only $2.33\times10^{-4}\%$ and
$2.14\times10^{-4}\%$, respectively, while for the global size
observables $W$, $H$, $A_{\rm sh}$, and $R_A$ they remain below
$1.2\times10^{-5}\%$. These variations are several orders of magnitude
smaller than the physical differences discussed in Sec.~\ref{subsec:running_M_Lambda_shadow}, showing that the reported shadow modifications are insensitive to the adopted
integration step at the relevant numerical accuracy. 

For the nearly polar configuration, $\theta_{\rm O}=0.2\,{\rm rad}$,
with the same values of $(a_*,\epsilon,\lambda,x_{\rm O})$,
we also compared the results obtained with
$h_\gamma=2.5\times10^{-4}$ and $h_\gamma=1.25\times10^{-4}$.
The $\theta_{\rm O}=0.2\,{\rm rad}$ results reported in
Table~\ref{tab:inclination_running_shadow} use the finer step.
Taking the finer-step result as the reference, halving
$h_\gamma$ changes $D_{cs}$ and $\delta_s$ by approximately
$1.89\times10^{-2}\%$ and $1.88\times10^{-2}\%$, respectively.
For the size observables $W$, $H$, $R_s$, $A_{\rm sh}$,
and $R_A$, the corresponding relative changes are all below
$6.4\times10^{-5}\%$.
These variations are well below the differences between the
running and Kerr--de Sitter geometries discussed in the main text.

Equatorial light rings are computed independently using
logarithmic root searches for the radial branches of the
circular-orbit equations. For the rotating configurations,
candidate solutions are retained when the absolute residuals
of both circular-orbit conditions are below $10^{-7}$.
A positive second radial derivative of the equatorial
potential selects the radially unstable candidates.

The numerical ray-tracing and shadow-extraction procedures were
validated in limits for which independent analytical results are
available. 
For the asymptotic benchmarks, we denote by $R_s^{\mathrm{B}}$
the reference-circle radius measured on the mass-normalized
Bardeen screen $(\alpha_{\mathrm{B}},\beta_{\mathrm{B}})$.
Its relation to the radius $R_s$ on the local gnomonic screen is
\begin{equation}
    R_s^{\mathrm{B}}
    \equiv
    \lim_{r_O/M\to\infty}
    \left(\frac{r_O}{M}\,R_s\right),
\end{equation}
where the limit is taken in the Kerr geometry,
$\ell=\Lambda_0=0$, at fixed $a/M$ and observer inclination
$\theta_O$. 
For example, as a quantitative benchmark,
for $a/M=0.784$, $\theta_{\rm O}=44.1^\circ$ and $\ell=\Lambda_0=0$
the asymptotic shadow calculation gives
$R_s^{\mathrm{B}}\simeq 5.10$,
$\delta_s\simeq0.050$, 
in agreement with the values quoted by
Hioki and Maeda~\cite{HiokiMaeda2009}.  (This comparison also tests the
conversion between the finite-distance ZAMO sky and the conventional
asymptotic impact-parameter screen).


\end{document}